\documentclass[11pt]{article}
\usepackage{amsmath}
\usepackage{a4wide}
\usepackage{tikz}
\usetikzlibrary{arrows,positioning,automata}
\RequirePackage[amsmath,hyperref,thmmarks]{ntheorem}
\theoremseparator{.}
\newtheorem{thm}{Theorem}
\newtheorem{lem}[thm]{Lemma}
\newtheorem{cor}[thm]{Corollary}

\theoremstyle{empty}
\renewtheorem{thm*}{Theorem}
\renewtheorem{lem*}{Lemma}
\renewtheorem{cor*}{Corollary}

\newenvironment{pf}[1][Proof]{\begin{trivlist}
\item[\hskip \labelsep {\bfseries #1} ] } {\end{trivlist}}

\renewcommand{\qed}{\nobreak \ifvmode \relax \else
     \ifdim\lastskip<1.5em \hskip-\lastskip
      \hskip1.5em plus0em minus0.5em \fi \nobreak
      \vrule height0.75em width0.5em depth0.25em\fi}

\title{Coloring Graphs having Few Colorings over Path Decompositions}

\author{Andreas Bj\"orklund\\Department of Computer Science, Lund University, Sweden\\andreas.bjorklund@yahoo.se} % Author name
\date{}

\begin{document}

\maketitle % Insert the title, author and date

\setlength\parindent{0pt} % Removes all indentation from paragraphs

\renewcommand{\labelenumi}{\alph{enumi}.} % Make numbering in the enumerate environment by letter rather than number (e.g. section 6)

\begin{abstract}
Lokshtanov, Marx, and Saurabh SODA 2011 proved that there is no $(k-\epsilon)^{\operatorname{pw}(G)}\operatorname{poly}(n)$ time algorithm for deciding if an $n$-vertex graph $G$ with pathwidth $\operatorname{pw}(G)$ admits a proper vertex coloring with $k$ colors unless the Strong Exponential Time Hypothesis (SETH) is false. 
We show here that nevertheless, when $k>\lfloor \Delta/2 \rfloor + 1$, where $\Delta$ is the maximum degree in the graph $G$, there is a better algorithm, at least when there are few colorings. We present a Monte Carlo algorithm that given a graph $G$ along with a path decomposition of $G$ with pathwidth $\operatorname{pw}(G)$ runs in $(\lfloor \Delta/2 \rfloor + 1)^{\operatorname{pw}(G)}\operatorname{poly}(n)s$ 
time, that distinguishes between $k$-colorable graphs having at most $s$ proper $k$-colorings and non-$k$-colorable graphs. 
We also show how to obtain a $k$-coloring in the same asymptotic running time.
Our algorithm avoids violating SETH for one since high degree vertices still cost too much and the mentioned hardness construction uses a lot of them. 

We exploit a new variation of the famous Alon--Tarsi theorem that has an algorithmic advantage over the original form. The original theorem shows a graph has an orientation with outdegree less than $k$ at every vertex, with a different number of odd and even Eulerian subgraphs only if the graph is $k$-colorable, but there is no known way of efficiently finding such an orientation. Our new form shows that if we instead count another difference of even and odd subgraphs meeting modular degree constraints at every vertex picked uniformly at random, we have a fair chance of getting a non-zero value if the graph has few $k$-colorings. Yet every non-$k$-colorable graph gives a zero difference, so a random set of constraints stands a good chance of being useful for separating the two cases.
\end{abstract}
%newpage

\section{Introduction}
One of the classical NP-hard problems on graphs is \emph{proper} vertex $k$-coloring~\cite{K72}: can you color the vertices from a palette of $k$ colors such that each pair of vertices connected by an edge are colored differently? The problem has numerous applications in both theory and practice, for instance to model resource allocation.

A well-known algorithmic technique to attack such a challenging task for many graphs is to use dynamic programming over a graph decomposition. In this paper we consider one of the most common ones, namely the path decomposition introduced in the seminal work on graph minors by Robertson and Seymour~\cite{RS83}. Lokshtanov, Marx, and Saurabh~\cite{LMS11} proved that there cannot exist a $(k-\epsilon)^{\operatorname{pw}(G)}\operatorname{poly}(n)$ time algorithm for deciding if an $n$-vertex graph $G$ with pathwidth $\operatorname{pw}(G)$ admits a proper vertex $k$-coloring for any $\epsilon>0$ unless the Strong Exponential Time Hypothesis is false. Actually, they state their result in terms of the more general concept treewidth, but the result holds as well for pathwidth as pointed out in their Theorem 6.1 \footnote{The Theorem says $(3-\epsilon)^{\operatorname{pw}(G)}$ but it is a misprint, it should be $(q-\epsilon)^{\operatorname{pw}(G)}$ in their notation.}.
The Strong Exponential Time Hypothesis (SETH)~\cite{IP99} says that $s_\infty=1$, where $s_k$ is the infimum of all real values $r$ for which there exists a $O(2^{rn})$ time algorithm that solves any $n$-variate $k$-\textsc{SAT} given in conjunctive normal form. In recent years many problems have been proven having known algorithms that are optimal under SETH, see e.g.~\cite{AV14, BI15, CKN13, LMS11}.

Indeed, for $k$-coloring nothing better than the natural $k^{\operatorname{pw}(G)}\operatorname{poly}(n)$ time algorithm that explicitly keeps track of all ways to color the presently active vertices is known for general graphs. However, the hardness construction from the result mentioned above uses many vertices of high degree and it would be interesting to understand to what extent this is necessary to enforce such strong lower bounds.

To this end we consider coloring bounded degree graphs with many colors. If the maximum degree in the graph is $\Delta$, it is trivial to find a $(\Delta+1)$-coloring by just coloring the vertices greedily in an arbitrary order. By Brook's theorem~\cite{B41}, one can also decide if there is a $\Delta$-coloring in polynomial time.
Reed~\cite{R99} goes even further and shows that for large enough $\Delta$, whenever there is no $\Delta$-clique, there is a $(\Delta-1)$-coloring. In general though, this is already a difficult coloring problem as it is NP-hard to $3$-color a graph of maximum degree $4$ (follows from taking the line graph of the construction in $\cite{H81}$).

We present in this paper an algorithm that is faster than the natural one when the number of colors $k\geq \lfloor \Delta/2\rfloor+1$. However, we also need that the number of $k$-colorings isn't too large as our algorithm gets \emph{slower} the more solutions there are. This counterintuitive behavior is symptomatic for the type of algorithm we use: we indirectly compute a fixed linear combination of all solutions and see if the result is non-zero. As the number of solutions increases, the number of ways that the solutions can annihilate each other also grows. Another example of such an algorithm (for directed Hamiltonian cycles) was recently given in \cite{BDH15}.

Still, already the class of uniquely $k$-colorable graphs is a rich and interesting one~\cite{X90}. One might suspect that it could be easier to find unique solutions.
However, there are parsimonious reductions from Satisfiability to $3$-coloring~\cite{B04} (up to permutations of the colors), and we know that unique Satisfiability isn't easier than the general case~\cite{C08}, so it cannot be too much easier. In particular, we still should expect it to take exponential time.

 \cite{HW08} uses ideas related to the present work to algebraically classify uniquely colorable graphs. The proposed means to solve for them though involve computations of Gr\"obner bases, which are known to be very slow in the worst case, and the paper does not discuss worst case computational efficiency. Our contribution here is to find a variation of the Alon-Tarsi theorem~\cite{AT92}, reusing the idea from~\cite{HW08} to look at the graph polynomial in points of powers of a primitive $k$:th root of unity, to get an efficient algorithm for solving a promise few $k$-coloring problem in bounded degree graphs. Our main theorem says
\begin{thm}
\label{thm: main}
For every undirected graph $G$, and path decomposition of $G$ of pathwidth $\operatorname{pw}(G)$,
there is a $(\lfloor \Delta/2 \rfloor+1)^{\operatorname{pw}(G)}\operatorname{poly}(n)s$ time Monte Carlo algorithm that outputs Yes with constant non-zero probability if $G$ is $k$-colorable but has at most $s$ proper $k$-colorings, and always outputs No if $G$ is non-$k$-colorable.
\end{thm}  

This means that for $k>\lfloor \Delta/2 \rfloor +1$ and small $s$ we improve exponentially over the natural $k^{\operatorname{pw}(G)}\operatorname{poly}(n)$ time algorithm. Still it does not violate the Strong Exponential Time Hypothesis lower bound from \cite{LMS11}, since that construction uses $\Omega(n/\log k)$ vertices of degree much larger than $2k$.

By a simple self-reduction argument, we can also obtain a witness coloring for the $k$-colorable graphs:
\begin{cor}
\label{cor}
For every $k$-colorable graph $G$ with $s$ proper $k$-colorings, and a path decomposition of $G$ of pathwidth $\operatorname{pw}(G)$, we can find a $k$-coloring in $(\lfloor \Delta/2 \rfloor+1)^{\operatorname{pw}(G)}\operatorname{poly}(n)s$ time, with constant non-zero probability.
\end{cor}

\subsection{The Alon--Tarsi theorem}
Let $G=(V,E)$ be an undirected graph,
and let $k$ be a positive integer. An \emph{orientation} of $G$ is a directed graph $D=(V,A)$ in which each edge $uv\in E$ is given an orientation, i.e. either $uv$ or $vu$ is in $A$ but not both. Denote by $\delta^+(A,v)$ and $\delta^-(A,v)$ the out- and indegree of the vertex $v$, respectively. 
A \emph{Eulerian subgraph} of $D$ is a subset $A'\subseteq A$ such that $\delta^+(A',v)=\delta^-(A',v)$ for all $v\in V$. Note that the notion of Eulerian here is somewhat non-standard as it does not require the subgraph to be connected.  The subgraph is \emph{even} if $|A'|$ is even and \emph{odd} otherwise. The theorem of Alon and Tarsi says
\begin{thm}{\cite{AT92}}
If there is an orientation $D=(V,A)$ of a graph $G$ such that $\delta^+(A,v)<k$ for all vertices $v\in V$ for some integer $k$, and the number of even and odd Eulerian subgraphs of $D$ differ, then $G$ is $k$-colorable.
\end{thm}
The theorem gives no promise in the other direction though, but Hefetz~\cite{H09} proved that if $G$ is uniquely $k$-colorable with a minimal number of edges then there also exist orientations meeting the criteria of the theorem. However, there are as far as we know no known ways of efficiently finding such an orientation for a general uniquely $k$-colorable graph and hence any successful algorithm for $k$-coloring based on computing the difference of the number of even and odd Eulerian subgraphs for a fixed orientation seems aloof. Still, this is what our algorithm does, albeit after relaxing Eulerian subgraphs to something broader.

\subsection{Our Approach}
We denote by $[k]$ the set $\{0,1,\cdots,k-1\}$. 
For a vector $w\in [k]^n$, which we also think of as a function $V\rightarrow [k]$, a \emph{$w$-mod-$k$ subgraph} is a subgraph with arc set $A'\subseteq A$ such that for all vertices $v\in V$, $\delta(A',v)\equiv w(v) (\mbox{ mod } k)$. 
Here, $\delta(A',v)=\delta^+(A',v)-\delta^-(A',v)$ is equal to the number of arcs in $A'$ outgoing from $v$ minus the arcs incoming to $v$.
 
Our algorithm is centered around a quantity $\kappa_{k,w}(A)$ that we define as the difference of the number of even $w$-mod-$k$ subgraphs and the number of odd ones.

%A \emph{partial} $k$-coloring is a function $c:V\rightarrow [k] \cup \{*\}$ such that for every edge $uv$ either $c(u)$ or $c(v)$ is $*$ or $c(u)\neq c(v)$. A \emph{completing} $k$-coloring is a coloring $c':V\rightarrow [k]$ that is consistent with $c$, i.e. $c(u)=c'(u)$ for all $u$ with $c(u)\neq *$. The \emph{color parts} are the sets $c'^{-1}(i)$ for $i\in [k]$.

Our main technical lemma that may be of independent combinatorial interest says
\begin{lem}
\label{lem: coldir}
Let an $n$-vertex graph $G$ and positive integer $k$ be given. If $G$ has $s$ proper $k$-colorings, then for any fixed orientation $A$ of the edges and a vector $w\in [k]^n$ chosen uniformly at random, it holds that
\begin{enumerate}
\item $\operatorname{P}(\kappa_{k,w}(A)\neq 0)=0$ if $s=0$,
\item $\operatorname{P}(\kappa_{k,w}(A)\neq 0)\geq s^{-1}$ if $s>0$.
\end{enumerate}
\end{lem}
Note that in this broad form the poor dependency on $s$ is best possible: the empty $n$-vertex graph has $s=k^n$ proper $k$-colorings but $\kappa_{k,w}(\emptyset)\neq 0$ only for $w=0$.

We will prove the Lemma in Section~\ref{sec: proof of lemma}. It immediately suggests an algorithm for separating $k$-colorable graphs with few colorings from non-$k$-colorable graphs that we will use to prove Theorem~\ref{thm: main}. The proof is in Section~\ref{sec: alg}. The algorithm is:

\vspace{5mm}
\noindent
\textbf{Decide-$k$-Colorable}
\begin{enumerate}
\item Pick any orientation $A$ of the edges.
\item Repeat $p(n,k)s$ times
\item \hspace{10mm} Pick a vector $w\in [k]^n$ uniformly at random.
\item \hspace{10mm} Compute $\kappa_{k,w}(A)$.
\item Output yes if $\kappa_{k,w}(A)\neq 0$ for any $w$, otherwise output no.
\end{enumerate}

In step d we compute $\kappa_{k,w}(A)$ over a path decomposition.
The key insight that makes this a faster algorithm than the natural one, is that we need to keep a much smaller state space when we count the $w$-mod-$k$ subgraphs than if we were to keep track of all colors explicitly. For a fixed orientation and decomposition, the edges are considered in a certain predetermined order during the execution of a path decomposition dynamic programming. Hence we only need to store states that we know will stand the chance to result in a $w$-mod-$k$ subgraph. To exemplify, say we are to count $w$-mod-$5$ subgraphs and a certain vertex $v$ has $w(v)=1$ and three incoming arcs and three outgoing arcs, and they are considered in order $++-+--$, where $+$ indicates an outgoing arc and $-$ an incoming. Now, when we have processed the first four of these, we only need to remember the partial solutions $A'\subseteq A$ that has $\delta(A',v)\in \{1,2,3\}$. It is possible to form partial solutions with $\delta(A',v)\in\{-1, 0\}$ as well, but the remaining two incoming arcs could never compensate for this imbalance to end up in a $\delta(A'',v)\equiv 1 (\operatorname{mod } 5)$ final state for some $A''\supseteq A'$.

%We also have
%\[
%\kappa_p(G')=\frac{1}{p^n}\sum_{v\in [m+1]^{p-1}} a_v \prod_{i=1}^{p-1} (1-\omega^{i})^{v_i},
%\]
%where $a_v$ for $v\in [m+1]^{p+1}$ counts the number of proper $p$-colorings $c$ of the graph $G'$ such that the number of arcs $uv$ with $c(u)-c(v)\equiv i (\operatorname{mod } p)$ is $v_i$.
\section{The Proof of Lemma~\ref{lem: coldir}}
\label{sec: proof of lemma}
Let $\omega$ be a primitive $k$:th root of unity over the complex numbers. Consider an undirected graph $G=(V,E)$ on $n$ vertices. For any directed graph $D=(V,B)$ where $uv\in E$ implies at least one of $uv$ and $vu$ to be in $B$, and $uv\in B$ implies $uv\in E$,
we define a graph function on any vertex coloring $c:V\rightarrow [k]$ as
\[
f_B(c)=\prod_{uv\in B} (1-\omega^{c(u)-c(v)}).
\]
Note that  $f_B(c)\neq 0$ if and only if $c$ is a proper $k$-coloring of $G$. Our previously defined difference $\kappa_{k,w}(B)$ of $w$-mod-$k$ subgraphs are related to $f_B$ through

\begin{lem}
\label{lem: kappa}
\begin{equation}
\label{eq: 1}
\kappa_{k,w}(B)=\frac{1}{k^n}\sum_{c\in [k]^n} \left(\prod_{v\in V} \omega^{-w(v)c(v)}\right)f_B(c).
\end{equation}
\end{lem}
\begin{pf}
We first observe that the right side of Eq.~\ref{eq: 1} can be rewritten as a summation over subgraphs by expanding the inner product, i.e.
\[
 \frac{1}{k^n}\sum_{c:V\rightarrow[k]} \prod_{v\in V}\omega^{-w(v)c(v)}\prod_{uv\in B} (1-\omega^{c(u)-c(v)}) = \frac{1}{k^n}\sum_{B'\subseteq B} (-1)^{|B'|}\sum_{c:V\rightarrow [k]} \prod_{v\in V} \omega^{(\delta(B',v)-w(v))c(v)}.
\]

Consider a fixed subgraph $B'\subseteq B$. It contributes the term
\[
\frac{1}{k^n} (-1)^{|B'|}\sum_{c:V\rightarrow [k]} \prod_{v\in V} \omega^{(\delta(B',v)-w(v))c(v)} .
\]

First observe that if there is a $u$ such that $k\!\!\not{|} (\delta(B',u)-w(u))$, then we can factor out $u$ from the expression to get
\[
\frac{1}{k^n} (-1)^{|B'|}\sum_{c_u\in [k]} \omega^{(\delta(B',u)-w(u))c_u} \sum_{c:V\setminus\{u\}\rightarrow [k]} \prod_{v\in V\setminus \{u\}} \omega^{(\delta(B',v)-w(v))c(v)}.
\]
Let $l=\delta(B',u)-w(u)$ and note that $(\sum_{c_u\in[k]} \omega^{lc_u}) (1-\omega^{l})=(1-\omega^{lk})=0$. Since $\omega$ is primitive and $k\!\!\not{|} l$ we have that  $(1-\omega^{l})$ is non-zero. We conclude that $\sum_{c_u\in[k]} \omega^{lc_u}=0$ and that such $B'$ contributes zero to the right hand expression of Eq.~\ref{eq: 1}.

Second note that when $k|(\delta(B',v)-w(v))$ for all $v\in V$, then the term $\prod_{v\in V} \omega^{(\delta(B',v)-w(v))c(v)}$ equals $1$ regardless of $c$ since $\omega$ is a $k$:th root of unity.
Hence such subgraphs $B'$ contributes the value $(-1)^{|B'|}$ after the division by the factor $k^n$ to the right hand expression of Eq.~\ref{eq: 1}.
We are left with $\sum_{w\operatorname{-mod-}k\operatorname{ subgraph } B'\subseteq B} (-1)^{|B'|}$ as claimed.
\qed
\end{pf}

In particular it follows from the above lemma that if $G$ is non-$k$-colorable, $\kappa_{k,w}(B)=0$ for every $w\in [k]^n$ because $f_B(c)\equiv 0$ in this case. Hence with $B=A$ we have proved item \emph{a} in Lemma~\ref{lem: coldir}.

To prove item \emph{b} of the Lemma, we will associate three directed graphs on $n$ vertices with $G$. First,
let $A$ be the fixed orientation of the edges in $E$ in the formulation of the Lemma. Second, let $\overline{A}$ be the reversal of $A$, i.e. for each arc $uv\in A$, $vu\in \overline{A}$. Finally, let $C=A \cup \overline{A}$.

\begin{lem}
\label{lem: bi-sum}
\[
\kappa_{k,0}(C)=\sum_{w\in [k]^n} \kappa_{k,w}(A)^2.
\]
\end{lem}
\begin{pf}
We first note that
\[
\kappa_{k,0}(C)=\sum_{w\in [k]^n} \kappa_{k,w}(A)\kappa_{k,-w}(\overline{A}).
\]
This is true because any even $0$-mod-$k$ graph in $C$ is either composed of an even $w$-mod-$k$ subgraph in $A$ and an even $(-w)$-mod-$k$ subgraph in $\overline{A}$ for some $w$, or is composed by two odd ones. Similarly, an odd $0$-mod-$k$ subgraph in $C$ is composed by an even-odd or odd-even pair in $A$ and $\overline{A}$ respectively for some $w$. Summing over all $w$ we count all subgraphs.
Next we note that $\kappa_{k,w}(A)=\kappa_{k,-w}(\overline{A})$ because $\delta(A,v)\equiv-\delta(\overline{A},v)(\mbox{ mod }k)$ for all $v\in V$.
\qed
\end{pf}

We will next use Lemma~\ref{lem: kappa} to bound $|\kappa_{k,0}(C)|$ and $|\kappa_{k,w}(A)|$. We have
\begin{lem}
\label{lem: bi-bound}
\[
|\kappa_{k,0}(C)|=\frac{1}{k^n}\sum_{c\in [k]^n} |f_A(c)|^2.
\]
\end{lem}
\begin{pf}
From Lemma~\ref{lem: kappa} we have
\[
\kappa_{k,0}(C)=\frac{1}{k^n}\sum_{c\in [k]^n} f_C(c).
\]
Since $f_C(c)=f_A(c)f_{\overline{A}}(c)$ and $f_{\overline{A}}(c)=\overline{f_{A}(c)}$, we have that $f_C(c)=|f_A(c)|^2$ and the Lemma follows. 
\qed
\end{pf}
\begin{lem}
\label{lem: one-bound}
\[
|\kappa_{k,w}(A)|\leq \frac{1}{k^n}\sum_{c\in [k]^n} |f_A(c)|.
\]
\end{lem}
\begin{pf}
From Lemma~\ref{lem: kappa} we have
\[
|\kappa_{k,w}(A)|=\frac{1}{k^n}\left|\sum_{c\in [k]^n}\left( \prod_{v\in V} \omega^{-w(v)c(v)}\right)f_A(c)\right|.
\]
Since
\[
\left|\sum_{c\in [k]^n} \left(\prod_{v\in V} \omega^{-w(v)c(v)}\right)f_A(c)\right| \leq \sum_{c\in [k]^n} \left(\prod_{v\in V} |\omega^{-w(v)c(v)}|\right)|f_A(c)|,
\]
and $|\omega^{-w(v)c(v)}|=1$ for every $w(v),c(v)$, the Lemma follows.
\qed
\end{pf}
Combining Lemmas~\ref{lem: bi-sum},~\ref{lem: bi-bound}, and~\ref{lem: one-bound}, we get
\[
\frac{1}{k^n}\sum_{c\in S} |f_A(c)|^2 \leq \sum_{w\in T} \left(\frac{1}{k^n}\sum_{c\in S} |f_A(c)|\right)^2.
\]
Here $S\subseteq [k]^n$ is the set of proper $k$-colorings and $T\subseteq [k]^n$ is the set of good $w$'s, i.e. $w\in T$ if and only if $\kappa_{k,w}(A)\neq 0$. By assuming that $s>0$ we can rewrite the above inequality as
\[
|T|\geq \frac{\frac{1}{k^n}\sum_{c\in S}|f_A(c)|^2}{\frac{1}{k^{2n}}(\sum_{c\in S} |f_A(c)|)^2}\geq \frac{k^n}{|S|},
\]
where the last inequality follows from Jensen's inequality. Dividing $|T|$ by $k^n$ gives the claimed probability bound in item \emph{b} of Lemma~\ref{lem: coldir}. This concludes the proof of our main Lemma.

\section{Details of the Algorithm}
\label{sec: alg}
We will prove Theorem~\ref{thm: main}. We will first describe how one can compute $\kappa_{k,w}(A)$ efficiently over a path decomposition and argue its correctness. Then we will prove Corollary~\ref{cor} by showing how one with polynomial overhead can obtain a witness $k$-coloring. 

\subsection{The Path Decomposition Algorithm}
\label{sec: pw}
Given a directed graph $H=(V,A)$ a \emph{path decomposition} of $H$ is a path graph $P=(U,F)$ where the vertices $U=\{u_1,\ldots,u_p\}$ represent subsets of $V$ called \emph{bags}, and the edges $F$ simply connect $u_i$ with $u_{i+1}$ for every $i<p$. Every vertex $v\in V$ is associated with an interval $I_v$ on $\{1,\ldots,p\}$ such that $v\in u_i$ iff $i\in I_v$. Furthermore, for each arc $ab\in A$, there exists an $i$ such that $\{a,b\}\subseteq u_i$. We set $r(ab)=i$ for the smallest such $i$. The \emph{pathwidth}, denoted $\operatorname{pw}(H)$, is the minimum over all path decompositions of $G$ of  $\operatorname{max}_i(|u_i|-1)$. In a \emph{nice} path decomposition, either a new vertex is added to $u_i$ to form $u_{i+1}$, in which case we call $u_{i+1}$ an \emph{introduce} bag, or a vertex is removed, in which case we call $u_{i+1}$ a \emph{forget} bag. We can assume w.l.o.g. that we have a nice path decomposition since it is straightforward to make any path decomposition nice by simply extending the path with enough bags.

We will see how one can compute $\kappa_{k,w}(A)$ for a fixed orientation $A$ over a path decomposition in step d in the algorithm \textbf{Decide-$k$-Colorable}.
We impose an ordering of the $m$ arcs, such that arc $a_i$ precede $a_{i+1}$ because $r(a_i)<r(a_{i+1})$  or $r(a_i)=r(a_{i+1})$ and $a_i$ is lexicographically before $a_{i+1}$.
We will loop over the arcs in the above order, virtually moving over the bags from $u_1$ to $u_p$ monotonically as necessary.
For arc $a_i$, we let $D_{i,v}$ for every vertex $v\in u_{r(a)}$ denote every possible modular degree difference $\delta(A',v) \mbox{ mod } k$ a vertex $v$ can have in a subgraph $A'\subseteq \{a_j:j\leq i\}$ such that there still are enough arcs in $\{a_j:j>i\}$ to form a subgraph $A'',A\supseteq A'' \supseteq A'$ with $k|(\delta(A'',v)-w(v))$. In particular, every $w$-mod-$k$ subgraph $A^*$ must have $\delta(A^*\cap \{a_j:j\leq i\},v)\in D_{i,v}$ for every $v\in u_{r(a_i)}$ and $i$.
\begin{lem}
\label{lem: delta bound}
For all $i$ and $v\in u_{r(a_i)}$,
\[
|D_{i,v}|\leq \lfloor \Delta/2 \rfloor + 1.
\]
\end{lem}
\begin{pf}
Let $D^{\mbox{\tiny before}}_{i,v}$ be the set of possible modular difference degrees $\delta(A',v) \mbox{ mod } k$ for any $A'\subseteq \{a_j:j\leq i\}$, and let $D^{\mbox{\tiny after}}_{i,v}$ be the set of possible negated difference degrees $(w(v)-\delta(A'',v)) \mbox{ mod } k$ for any $A''\subseteq \{a_j:j>i\}$. Observe that $D_{i,v}=D^{\mbox{\tiny before}}_{i,v} \cap D^{\mbox{\tiny after}}_{i,v}$. If the number of arcs incident to $v$ in $\{a_j:j\leq i\}$ is $d^{\mbox{\tiny before}}_v$, and the ones in $\{a_j:j>i\}$ is $d^{\mbox{\tiny after}}_v$, we have that $|D^{\mbox{\tiny before}}_{i,v}|\leq d^{\mbox{\tiny before}}+1$ and $|D^{\mbox{\tiny after}}_{i,v}|\leq d^{\mbox{\tiny after}}+1$ . Since $\Delta\geq d^{\mbox{\tiny before}}_v+d^{\mbox{\tiny after}}_v$, the bound follows. \qed
\end{pf}

Our algorithm tabulates for each possible modular degree difference in $D_{i,v}$ for each $v\in u_{r(a_i)}$, the difference of the number of even and odd subgraphs in $\{a_j:j\leq i\}$ matching the degree constraints on those vertices, while having modular degree difference equal to $w(v)$ on every vertex $v$ that are forgotten by the algorithm, i.e. vertices $v$ that were abandoned in a forget bag $u_j$ for $j<r(a_i)$.
That is, the complete state is described by a function $s_i:D_{i,v_1}\times\cdots \times D_{i,v_l}\rightarrow \mathbf{Z}$, where $\{v_1,\ldots,v_l\}=u_{r(a_i)}$, and where $s_i$ for a specific difference degree vector holds the above difference of the number of even and odd subgraphs.
It is easy to compute $s_{i+1}$ from $s_i$, since each point in $s_{i+1}$ depends on at most two points of $s_i$ (either we use the arc $a_{i+1}$ in our partial $w$-mod-$k$ subgraph, or we do not). To compute the new value we just subtract the two old in reversed order, i.e. with an abuse of notation $s_{i+1}(d)=s_{i}(d)-s_{i}(d - a_{i+1})$. We initialize $s_0$ to all-zero except for $s_0(0)=1$ since the empty subgraph is an even $0$-mod-$k$ subgraph. We store $s_i$ in an array sorted after the lexicographically order on the (Cartesian product) keys which allows for quick access when we construct $s_{i+1}$. We only need to precompute $D_{i,v}$ for all $i$ and $v$ which is easily done in polynomial time, in order to see what modular degree differences to consider for $s_{i+1}$ and what is stored in the previous function table $s_i$. Once we have computed $s_{m}$, we can read off $\kappa_{k,w}(A)$ from $s_m(w)$.
 
\subsection{Runtime and Correctness Analysis}
We finish the proof of Theorem~\ref{thm: main}. 
It follows from the bound in Lemma~\ref{lem: delta bound} that step d of the algorithm takes $(\lfloor \Delta/2 \rfloor+1)^{\operatorname{pw}(G)}\operatorname{poly}(n)$ time.  It is executed $p(n)s$ times so that we in expectation sample $p(n,k)$ good $w$'s for which $\kappa_{k,w}(A)\neq 0$ for $k$-colorable graphs having at most $s$ proper $k$-colorings, as seen from item \emph{b} in Lemma~\ref{lem: coldir}. From Markov's inequality, the probability of false negatives is at most $\frac{1}{p(n,k)}$. Thus already $p(n,k)=2$ will do. From item \emph{a} in Lemma~\ref{lem: coldir} the probability of false positives is zero.
 
\subsection{Coloring a Graph}
We proceed with the proof of Corollary~\ref{cor}.
The idea for recovering a witness $k$-coloring is to use self-reduction, i.e. to use algorithm \textbf{Decide-$k$-Colorable} many times for several graphs obtained by modifying $G$ so as to gradually learn more and more of the vertices' colors. We will learn a coloring one color a time. That is, we will find a maximal subset of the vertices that can be colored in one color so that the remaining graph can be colored by the remaining $k-1$ colors.

For every subset $S\subseteq V$ of the vertices that form an independent set in $G$, we let $G_S$ be the graph obtained by collapsing all vertices in $S$ into a single supervertex $v_S$ that retain all edges that goes to $S$ in the original graph. Note that the number of colorings cannot increase by the contraction, and the degree is only increased for the newly formed vertex $v_S$, the rest of the graph is intact. Also note that the path width increases by at most one, since removing $v_S$ from the graph leaves a subgraph of $G$. We use algorithm \textbf{Decide-$k$-Colorable} on $G_S$ as a subroutine and note that the runtime is at most a factor $k$ larger than it was on $G$ due to the supervertex $v_S$ that now may be in all bags and we need to keep track of all modular degree differences for this vertex. Our algorithm for extracting a color class is:

\vspace{5mm}
\noindent
\textbf{Find-Maximal-Color-Class}
\begin{enumerate}
\item Let $S={v_1}$.
\item For every vertex $u\in V\setminus{v_1}$,
\item \hspace{10mm} If $S\cup u$ is an independent set and \textbf{Decide-$k$-Colorable($G_{S\cup \{u\}}$)} returns Yes,
\item \hspace{20mm} Let $S=S\cup \{u\}$.
\item Return $S$.
\end{enumerate}
With $p(n,k)=2nk$ we get the true verdict on all queried graphs with probability at least $1-\frac{1}{2k}$.
Once we have found a maximal color class $S$, we can continue coloring the induced subgraph $G[V\setminus S]$ which is $(k-1)$-colorable by extracting a second color class and so on.  We note that neither the maximum degree, pathwidth, or number of proper colorings can increase in an induced subgraph. In particular, a path decomposition for $G[V\setminus S]$ of pathwidth at most $\operatorname{pw}(G)$ can be readily obtained from the given path decomposition for $G$ by just omitting the removed vertices and edges. By the union bound, with probability at least $\frac{1}{2}$ we correctly extract all $k$ color classes.
 
\section{Improvements and Limitations}
The striking dependence on $s$ in Theorem~\ref{thm: main} is at least in part due to the poor uniform sampling 
employed. For many $w$'s  $\kappa_{k,w}(A)$ is zero for a trivial reason, namely that there are no $w$-mod-$k$ subgraphs at all. A better idea would be to try to sample uniformly over all $w$'s that has at least one $w$-mod-$k$-subgraph. We don't know how to do that, but it is probably still better to sample non-uniformly over this subset of good $w$'s by uniformly picking a subgraph of $G$, and letting $w$ be given by the degree differences of that subgraph. However, we can give an example of graphs where even a uniform sampling over the attainable $w$'s will require running time that grows with $s$. Our construction is very simple, we just consider the graph consisting of $n/3$ disjoint triangles.
Let $A$ be an orientation such that each vertex has one incoming and one outgoing arc. Then, the number of attainable $w$'s is $7^{n/3}$ since every non-empty subset of the three arcs in a triangle gives a unique modular degree difference.
The number of $w$'s that give a non-zero $\kappa_{3,w}(A)$ is just $6^{n/3}$ which can be seen by inspecting each of the $7$ attainable $w$'s for a triangle, and noting that $\kappa_{k,w}(A' \cup A")=\kappa_{k,w'}(A')\kappa_{k,w"}(A")$ for vertex-disjoint arc subsets $A'$ and $A"$. The number of $3$-colorings $s$ is also $6^{n/3}$, so with probability $s^{-0.086}$ we pick a $w$ that has $\kappa_{3,w}(A)\neq 0$. While being a great improvement over $s^{-1}$, it still demonstrates a severe limitation of the technique presented in this paper when there are many solutions.

\section*{Acknowledgments}
I thank several anonymous reviewers for comments on an earlier version of the paper.
This research was supported in part by the Swedish Research Council grant VR 2012-4730 Exact Exponential Algorithms.

\end{document}